\documentclass[aps,prl,twocolumn,superscriptaddress,nofootinbib]{revtex4-2}

\usepackage{graphicx}
\usepackage{amsmath}
\usepackage{xcolor}
\usepackage[normalem]{ulem}

\newcommand{\prsec}[1]{\textit{#1}.}
\begin{document}

\title{Active-Learning Inspired \textit{Ab Initio} Theory-Experiment Loop Approach for Management of Material Defects: Application to Superconducting Qubits}

\author{Sarvesh Chaudhari}
\email{sc2923@cornell.edu}
\affiliation{Department of Physics, Cornell University}

\author{Crist\'obal M\'endez}
\affiliation{School of Applied and Engineering Physics, Cornell University}

\author{Rushil Choudhary}
\affiliation{Department of Physics, Cornell University}

\author{Tathagata Banerjee}
\affiliation{School of Applied and Engineering Physics, Cornell University}

\author{Maciej W. Olszewski}
\affiliation{Department of Physics, Cornell University}

\author{Jadrien T. Paustian}
\affiliation{Department of Physics, Syracuse University}

\author{Jaehong Choi}
\affiliation{School of Applied and Engineering Physics, Cornell University}

\author{Zhaslan Baraissov}
\affiliation{School of Applied and Engineering Physics, Cornell University}

\author{Raul Hernandez}
\affiliation{Department of Electrical Engineering and Computer Science, Massachusetts Institute of Technology}

\author{David A. Muller}
\affiliation{School of Applied and Engineering Physics, Cornell University}

\author{B. L. T. Plourde}
\affiliation{Department of Physics, Syracuse University}

\author{Gregory D. Fuchs}
\affiliation{School of Applied and Engineering Physics, Cornell University}

\author{Valla Fatemi}
\affiliation{School of Applied and Engineering Physics, Cornell University}

\author{Tom\'as A. Arias}
\affiliation{Department of Physics, Cornell University}

\begin{abstract}
Surface oxides are associated with two-level systems (TLSs) that degrade the performance of niobium-based superconducting quantum computing devices. To address this, we introduce a predictive framework for selecting metal capping layers that inhibit niobium oxide formation. Using DFT-calculated oxygen interstitial and vacancy energies as thermodynamic descriptors, we train a logistic regression model on a limited set of experimental outcomes to successfully predict the likelihood of oxide formation beneath different capping materials. This approach identifies Zr, Hf, and Ta as effective diffusion barriers. Our analysis further reveals that the oxide formation energy per oxygen atom serves as an excellent standalone descriptor for predicting barrier performance. By combining this new descriptor with lattice mismatch as a secondary criterion to promote structurally coherent interfaces, we identify Zr, Ta, and Sc as especially promising candidates. This closed-loop strategy integrates first-principles theory, machine learning, and limited experimental data to enable rational design of next-generation materials.
\end{abstract}

\maketitle

\prsec{\label{sec:Introduction} Introduction}
The rational design of materials has played a pivotal role in advancing various technologies by accelerating the development of materials with tailored properties. This approach has been extensively applied in fields ranging from fuel cell catalysis \cite{fuelcell}, energy storage materials \cite{energystorage}, and photonic structures \cite{photonic} to high performance polymers \cite{TARDIO2024112310}. These advances have been enabled by computational tools, high-throughput screening, and increasingly, machine learning, which allow for the optimization of bulk properties including bandgaps \cite{Zhu2020}, thermal conductivity \cite{Zhou2021}, and electrochemical activity \cite{Guan2025}. As device dimensions shrink and interfaces dominate performance, local atomic-scale processes become as critical as bulk properties. Defect energetics are therefore widely studied using both first-principles and data-driven approaches~\cite{deml2015intrinsic,kumagai2020insights,kiyohara2025charged,tran2018active,wang2020stacking,chaudhary2017stacking,zhang2016defect}, yet they are rarely employed as explicit forward-design variables in theory--experiment closed-loop frameworks linked to device-level failure mechanisms.

This design perspective is particularly critical in quantum computing \cite{deLeon2021, brown2021}, specifically in superconducting quantum devices, where Josephson junctions (JJs) are formed by sandwiching thin insulating or oxidized layers between superconducting electrodes. In such systems, oxygen diffusion from adjacent capping layers or the ambient environment can lead to the formation of unwanted oxides at the interfaces, most notably niobium oxide in niobium-based superconducting JJs \cite{romanenko2020three, murthy2022}. These oxides host two-level systems (TLSs) that couple to the electric field of the qubit, acting as parasitic resonators that degrade coherence times and limit device performance \cite{Mueller2019TLS, bafia2024oxygen}.

Suppressing the formation of such oxides is therefore a key challenge in improving Nb JJ qubit performance. Various strategies in this regard have been explored, including high-temperature heat treatment and etching, which reduce TLS losses \cite{romanenko2020three, kalboussi2025, alghadeer2024, altoe2022}. Also promising from a materials-design standpoint is encapsulation of the superconducting metal with an oxygen barrier surface, which has demonstrated encouraging experimental performance \cite{bal2024}. However, barrier-material selection and optimization remain largely empirical, with limited predictive guidance.

\begin{figure}[t]
    \centering
    \includegraphics[width=\columnwidth]{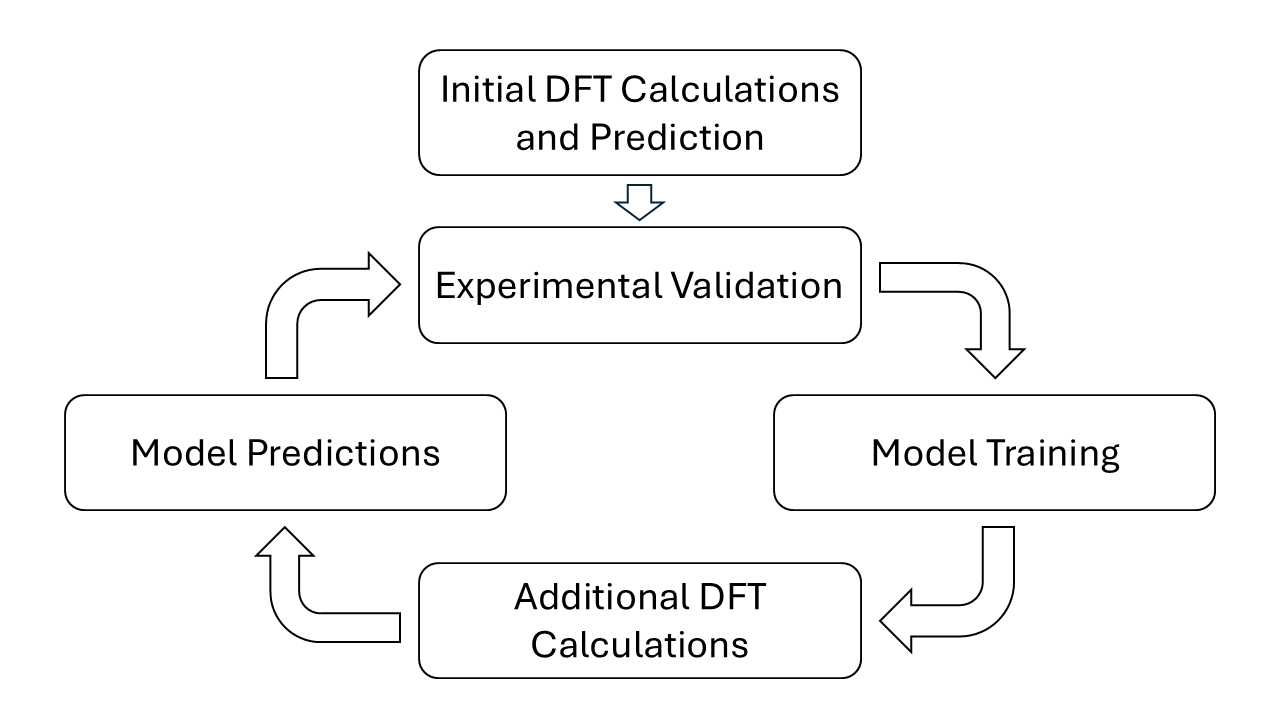}
    \caption{Schematic overview of the theory–experiment active-learning workflow used in this work.}
    \label{fig:workflow}
\end{figure}

Within the Materials Genome framework, enabling predictive design of oxidation-resistant barriers naturally invites active-learning–inspired materials-selection strategies, including surrogate optimization~\cite{hase2018phoenics,tran2018active,oftelie2018active,lookman2019active,dunn2019rocketsled,montoya2020autonomous} and feasibility-based screening~\cite{kishio2013strategic,takahashi2022adaptive}. However, such approaches are most commonly deployed in high-throughput, data-rich settings and offer limited physical interpretability, motivating the development of interpretable active-learning frameworks suited to sparse-data materials-design problems.

In this work, we introduce a predictive framework that leverages defect energies calculated from first principles to guide the selection of oxidation-resistant barrier materials for niobium-based superconducting qubits. By quantifying key microscopic energetics, vacancy formation energies in metal oxides and oxygen interstitial energies in metals, we construct an interpretable logistic-regression model that captures the likelihood of interfacial oxide formation across candidate barriers. The resulting descriptor-space map distinguishes effective from ineffective oxygen barriers and enables prediction of new viable materials. 
The model operates in a closed-loop workflow in which first principles energetics define the initial predictions and guide early experiments. Experimental outcomes then train the model, which helps direct targeted DFT calculations to update the descriptor space, enabling subsequent cycles of experiments, DFT calculations, and model refinement (Fig.~\ref{fig:workflow}). More broadly, this framework illustrates how theory and experiment can be integrated in sparse-data settings through an interpretable, active-learning–inspired implementation of the Materials Genome design loop.

\prsec{Computational details}
All \emph{ab initio} density–functional theory (DFT) calculations were performed using the open-source plane-wave code \textsc{JDFTx}~\cite{payne1992iterative,sundararaman2017jdftx} with ultrasoft GBRV pseudopotentials~\cite{GARRITY2014446} and the PBEsol exchange-correlation functional~\cite{perdew2008restoring}. Computational parameters, structural models, convergence criteria, and defect-supercell specifications are provided in the Supplemental Material~\cite{SupplementalMaterial} (see also reference \cite{Jain2013} therein).

\prsec{Experimental details} Niobium thin films were deposited on high-resistivity Si(100) substrates by Ar magnetron sputtering and patterned using standard photolithography and reactive-ion etching. Samples received post-fabrication surface treatments prior to characterization. X-ray photoelectron spectroscopy (XPS) measurements were performed to quantify interfacial oxide formation and to classify diffusion-barrier performance. Fabrication protocols, surface treatments, measurement conditions, and spectral fitting procedures are provided in the Supplemental Material~\cite{SupplementalMaterial} (see also references \cite{olszewski2025lowlossnbsisuperconducting,FAIRLEY2021100112,Banerjee} therein).

\prsec{\label{sec:lr}Logistic regression}
Logistic regression is a supervised machine learning method well suited for modeling binary classification problems using continuous-valued descriptors \cite{abd18569-0124-3b1e-9ca9-67a6fd857a26}. In the context of this study, it is used to predict the probability that niobium oxide will form at an interface, based on defect energies of candidate barrier materials. This model is integrated into an active learning framework, where predictions are iteratively refined through a theory-experiment loop: candidate materials are recommended by the model, tested experimentally, and the resulting data are used to further train the model.

The model estimates the probability $P(x_i | \beta_i)$ of oxide formation as a function of input features $x_i = \{x_1, x_2, ..., x_n\}$ (\emph{e.g.} defect formation energies) and fitting parameters $\beta_i = \{\beta_0, \beta_1, \beta_2, ..., \beta_n\}$. The prediction is made using a sigmoid function, which maps the weighted sum of descriptors to a probability between 0 and 1, and takes the following form:
\begin{equation}
    \label{sigmoid}
    P(x_i | \beta_i) = \frac{1}{1 + \exp \left[ - \left( \beta_0 + \beta_1 x_1 + ... + \beta_n x_n \right) \right]}
\end{equation}
where the coefficients $\beta_i$ are learned by minimizing the cross entropy between the predicted and experimental probabilities from the training data set. The cross entropy is defined as:
\begin{equation}
    \label{cross_entropy}
    H = - \sum_{i=1}^{N} \left[ y_i \log P(x_i | \beta_i) + (1 - y_i) \log (1 - P(x_i | \beta_i)) \right]
\end{equation}
where $y_i \in [0, 1]$ denotes the experimental probability of oxide formation for sample $i$, and $N$ is the total number of training samples.

Because the model predicts the probability of niobium oxide formation, it must account for the inherent uncertainty in the experimental training data. To incorporate this, binary observations of whether niobium oxide formed or not were converted into soft labels that reflect the confidence level in each experimental outcome. Metals that were observed to permit oxide formation were assigned experimental probabilities around 0.9, based on the likelihood that the observed oxide originated from diffusion processes rather than fabrication defects. If sample quality was independently verified through direct inspection, an experimental probability of 1 was assigned. Conversely, materials that prevented oxide formation were assigned experimental probabilities of 0. This assignment reflects the fact that, regardless of uncertainties in fabrication conditions, the metal successfully blocked oxide formation. The logistic regression model was then trained by minimizing the cross-entropy between the predicted probabilities and these experimental probabilities, thereby incorporating experimental uncertainty into the learning process.

\prsec{\label{sec:Initial Descriptors} Defect descriptors}
To predict niobium oxide formation beneath metal capping layers, relevant thermodynamic descriptors were identified based on the microscopic processes that govern oxygen diffusion toward the niobium substrate. These processes are:

\begin{enumerate}
    \item Metal Oxide to Metal transfers: In this process, an oxygen atom leaves the metal oxide leaving behind a vacancy and joins the metal in an interstitial site. The change in energy associated with this is
    \begin{equation}
        \label{MOtM}
        \Delta E_{\mathrm{MO \rightarrow M}} = E(V_\mathrm{O}(\mathrm{M}_n \mathrm{O}_m)) + E(\mathrm{O}_I(\mathrm{M})),
    \end{equation}
    where $E(V_\mathrm{O}(\mathrm{M}_n \mathrm{O}_m))$ is the energy to form an oxygen vacancy in the oxide $\mathrm{M}_n \mathrm{O}_m$, and $E(\mathrm{O}_I(\mathrm{M}))$ is the energy to add an oxygen interstitial into a metal $\mathrm{M}$. The details of the reference energies used for determining interstitial and vacancy energies are described below.
    \item Metal to Niobium transfers: In this process, an oxygen atom leaves an interstitial site in the metal and joins the niobium in an interstitial site. This process is important to begin the formation of the niobium oxide. The change in energy associated with this is
    \begin{equation}
        \label{MtNb}
        \Delta E_{\mathrm{M \rightarrow Nb}} = -E(\mathrm{O}_I(\mathrm{M})) + E(\mathrm{O}_I(\mathrm{Nb})).
    \end{equation}
    \item Metal to Niobium Oxide transfers: In this process, an oxygen atom leaves an interstitial site in the metal and fills in a vacancy in the niobium oxide. This process is important for completing the niobium oxide. The change in energy associated with this process is
    \begin{equation}
        \label{MtNbOx}
        \Delta E_{\mathrm{M \rightarrow NbOx}} = -E(\mathrm{O}_I(\mathrm{M})) - E(V_\mathrm{O}(\mathrm{Nb}_n \mathrm{O}_m)).
    \end{equation}
\end{enumerate}

The quantities that vary among different capping metals in Equations \ref{MOtM}, \ref{MtNb}, and \ref{MtNbOx} are $E(V_\mathrm{O}(\mathrm{M}_n \mathrm{O}_m))$ and $E(\mathrm{O}_I(\mathrm{M}))$. These are the defect energies that will be used as input features in the logistic regression model.

To evaluate our input features \emph{ab initio}, we compute the oxygen interstitial energy (with O$_2$ as the reference for the chemical potential of O) as
\begin{equation}
    E(\mathrm{O}_I(\mathrm{M})) = E(\mathrm{M}_N\mathrm{O}) - \left[ E(\mathrm{M}_N) + \frac{1}{2}E(\mathrm{O}_2)\right],
\end{equation}
where $E(\mathrm{M}_N\mathrm{O})$ is the energy of a computational supercell with $N$ atoms of metal M and one O interstitial, $E(\mathrm{M}_N)$ is the energy of a supercell with $N$ atoms of metal M, and $E(\mathrm{O}_2)$ is the energy of an oxygen molecule, which ultimately cancels out in Equations \ref{MOtM}, \ref{MtNb}, and \ref{MtNbOx}. Similarly, the oxygen vacancy energy is computed as
\begin{align}
    E(V_\mathrm{O}(\mathrm{M}_n \mathrm{O}_m)) = &\left[ E(\mathrm{M}_{Nn}\mathrm{O}_{Nm-1}) + \frac{1}{2} E(\mathrm{O}_2) \right] \notag \\
    &- E(\mathrm{M}_{Nn}\mathrm{O}_{Nm})
\end{align}
where $E(\mathrm{M}_{Nn}\mathrm{O}_{Nm-1})$ is the energy of a supercell with $N$ unit cells of the oxide $\mathrm{M}_n\mathrm{O}_m$ minus one O atom, and $E(\mathrm{M}_{Nn}\mathrm{O}_{Nm})$ is the energy of a supercell with $N$ unit cells of the oxide $\mathrm{M}_n\mathrm{O}_m$. Because our objective is to construct comparative descriptors rather than reproduce experimental defect formation energies, we evaluate neutral defect energetics in the expressions above for consistency and transferability. Charged-defect calculations require specification of an electron chemical potential that depends sensitively on interface and device conditions and thus introduce ambiguities. Although defect charge states and operating environments may vary, neutral energetics remain positively correlated with charged-defect energetics and therefore serve as informative, transferable descriptors for diffusion-barrier performance.

\begin{figure}
    \centering
    \includegraphics[width=\columnwidth]{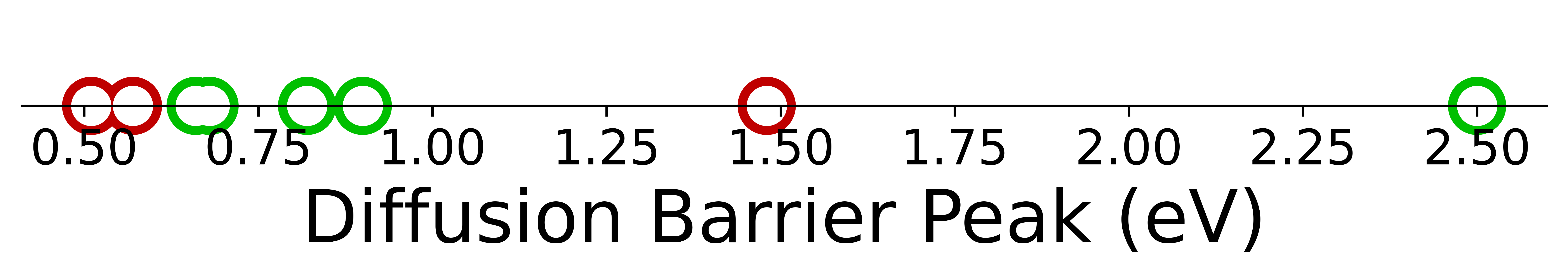}
    \caption{Diffusion barrier performance versus diffusion barrier energy: barrier energy (position along horizontal axis), performance (green=good, red=poor) for Au, Pd, W, Ta, Mo, Al, Pt, and Zr from left to right. Data show no clear correlation between barrier energies and performance.}
    \label{fig:diff}
\end{figure}

To evaluate whether kinetic limitations might influence oxide formation, we also calculated diffusion barriers for oxygen hopping between interstitial sites in each barrier metal. However, as shown in Figure~\ref{fig:diff}, there is no correlation between the computed diffusion barriers and the experimentally observed ability of metals to suppress niobium oxide formation. This strongly suggests that thermodynamic factors, not kinetics, control the outcome.

To reinforce this conclusion, we estimated oxygen diffusion times across a 5-nm capping layer using Arrhenius diffusion models (see Supplemental Material \cite{SupplementalMaterial}). Traversal occurs on sub-second to minute timescales for representative barriers (0.5--0.8~eV), indicating that oxygen transport is generally rapid within experimental timeframes. However, Figure~\ref{fig:diff} shows that low-barrier materials can still suppress oxide formation, whereas the highest-barrier case ($\sim$1.5~eV) permits oxidation despite an estimated diffusion time $\sim 10^{12}$ times longer than for a 0.8~eV barrier. This implies that oxygen can access alternative pathways (\emph{e.g.}, grain boundaries) to traverse the barrier. Combined with the lack of correlation between diffusion barriers and observed oxide suppression, these findings indicate primarily thermodynamic rather than kinetic control of oxide formation. We therefore use only the interstitial and vacancy formation energies as descriptors in our model.

\begin{figure}
    \centering
    \includegraphics[width=\columnwidth]{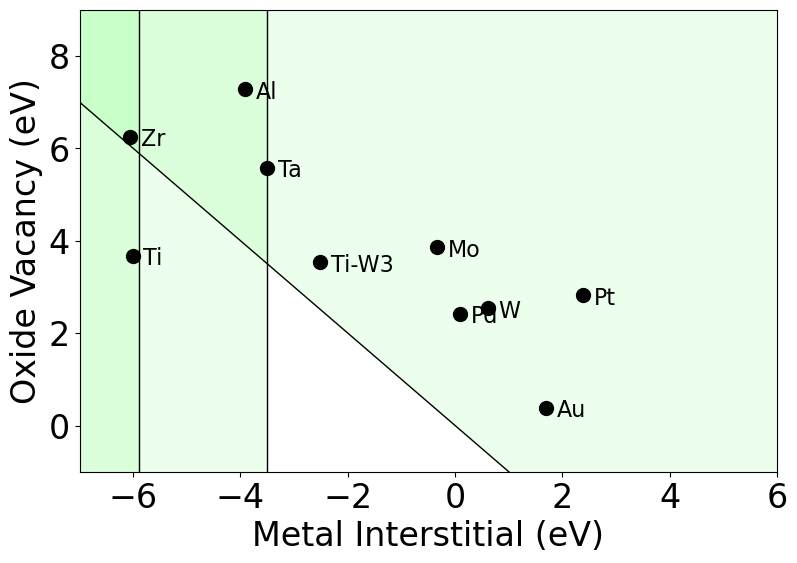}
    \caption{Oxide vacancy energy versus metal interstitial energy for a set of metals. Darker green regions indicate increasingly favorable thermodynamic barriers to oxygen diffusion.}
    \label{fig:phase_no_exp}
\end{figure}

\prsec{Descriptor map}
Figure~\ref{fig:phase_no_exp} shows a descriptor map of oxygen vacancy energy versus metal interstitial energy for the initial set of candidate barrier metals. This materials set was selected to span established junction constituents and technologically relevant barrier chemistries, including noble metals resistant to oxidation (Au, Pt, Pd); Al, the widely used Al$_2$O$_3$ tunnel-barrier constituent in Nb-based junctions~\cite{10.1109/19.982937}; refractory and Nb-adjacent transition metals (Ta, Ti, Zr, W, Mo), selected for chemical similarity, lattice compatibility, and strong oxygen affinity; and TiW$_x$, a refractory diffusion-barrier alloy widely employed in thin-film metallization stacks~\cite{10.1063/5.0048304}.
\ \\
Our initial processes in Equations \ref{MOtM}, \ref{MtNb}, and \ref{MtNbOx} represent the sequential microscopic steps necessary to form niobium oxide. For a metal to be a good diffusion barrier, we would like each of these processes to be thermodynamically unfavorable ($\Delta E > 0$). Each of these conditions corresponds to a linear inequality in our two-dimensional descriptor space, forming a half-plane boundary. In Figure~\ref{fig:phase_no_exp}, the shaded green regions indicate the zones where each individual process is suppressed. Where these regions overlap, the shading becomes darker, indicating that multiple (or all) oxygen transfer pathways are energetically blocked. Metals located in these darker regions are therefore expected to serve as more effective diffusion barriers.

Most metals explored lie above the diagonal boundary corresponding to Equation~\ref{MOtM}, which prevents oxygen from leaching out of the surface oxide and entering the barrier metal. The vertical boundary near $E(\mathrm{O}_I(\mathrm{M})) \sim -3.5$~eV (from Equation~\ref{MtNb}) excludes metals that would readily transfer interstitial oxygen into metallic niobium as an interstitial. A stricter boundary near $E(\mathrm{O}_I(\mathrm{M})) \sim -6$~eV (from Equation~\ref{MtNbOx}) further excludes metals that would facilitate oxygen interstitial transfer into pre-existing niobium oxide vacancies. Both of these vertical cuts suggest that moving leftward in the plot corresponds to increasing thermodynamic resistance to oxide formation. Together, these three conditions indicate that materials in the upper-left quadrant of the descriptor space are most favorable, while those in the lower-right are least favorable. 

At this level of analysis, zirconium stands out as satisfying all three criteria and therefore emerges as a particularly promising barrier material. Even if fully oxidized, Eqs.~\ref{MOtM} and~\ref{MtNb} indicate that zirconium should maintain a chemically sharp interface with niobium, making zirconium oxide a compelling insulating candidate for Josephson junctions. This thermodynamic picture is consistent with prior studies from the superconducting radio-frequency community documenting zirconium surface segregation and preferential oxidation in niobium–zirconium alloys~\cite{10.7298/mc9b-ty97,osti_4268078,10.1103/PhysRevApplied.20.014064,10.1002/aelm.202300151}, as well as direct observations of passivating ZrO$_2$ formation on niobium surfaces~\cite{10.18429/JACoW-SRF2023-TUPTB004}. In contrast, the present work addresses the defect energetics and interfacial stability governing diffusion-barrier behavior. Building on the thermodynamic predictions presented here, recent experimental studies of Nb/ZrO$_x$/Nb junctions have reported exceptionally abrupt interfaces, consistent with the predicted oxygen segregation behavior~\cite{10.1063/5.0296881}.

\begin{figure}[h]
    \centering
    \begin{center}
        \includegraphics[width=\columnwidth]{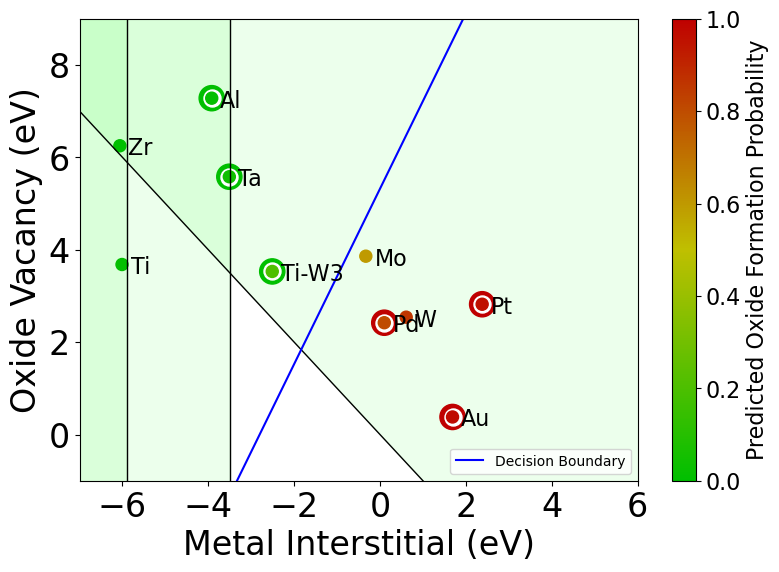}
        {\bf\large (a)}
        \ \\
        \ \\
        \includegraphics[width=\columnwidth]{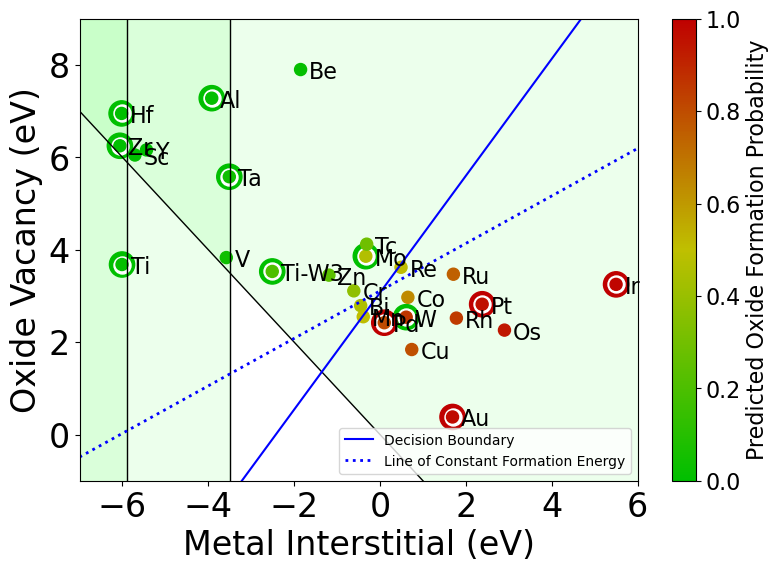}
        {\bf\large (b)}
    \end{center}
    \caption{Oxygen vacancy energy vs. metal interstitial energy across logistic regression iterations: first iteration (a), final iteration (b). Experimentally tested metals are outlined in green (success) or red (failure). Point colors reflect predicted oxide formation probabilities. The solid blue line shows the model’s decision boundary; the dashed line indicates constant oxide formation energy.}
    \label{fig:phase}
\end{figure}

\prsec{Logistic theory-experiment loop}
We note that the thermodynamic criteria discussed in the previous section, while physically motivated, may be too strict. Each of the three processes was treated independently, and the overall conclusion required all of them to be simultaneously unfavorable. However, in practice, preventing even a single step in the oxygen transfer pathway may be sufficient to block niobium oxide formation. Furthermore, the prior analysis does not account for competing effects (such as kinetics) that may impact the outcome over finite time scales. To move beyond a binary thermodynamic filtering, we adopt a data-driven approach based on logistic regression, which provides a probabilistic model trained directly on experimental results.

Our initial experimental dataset corresponds to the circled points in Fig.~\ref{fig:phase}a, selected as a representative subset of the \emph{ab initio} set to broadly sample the descriptor space while also reflecting practical considerations of fabrication accessibility and turnaround. Experimental outcomes for Al, Ta, TiW$_3$, Pd, Pt, and Au are indicated by green or red circles outlining the data points, where green and red denote whether niobium oxide formation was observed to be suppressed or not. These data confirmed the broad trends predicted by the descriptor space, with successful barrier metals appearing toward the upper left and unsuccessful metals toward the lower right. We then used this data to train the first iteration of our logistic regression model as described above in \emph{Logistic regression}, initiating the theory-experiment loop depicted in Figure~\ref{fig:workflow}. The predicted probabilities are encoded in the figure as the color of the inner solid circles, ranging from green ($P=0$) to red ($P=1$).

Metals were then prioritized for further testing based on their predicted probability of success. After each new set of experiments, the results were fed back into the model, which was retrained to improve its predictive accuracy. This created a closed theory-experiment loop that progressed over several iterations. The materials search space encompassed 27 candidate metals relevant to niobium-based junctions, selected to span transition-metal rows, d-band fillings, and selected s- and p-block native-oxide chemistries. Aluminum oxide provides the canonical barrier reference, while Be-, Zn-, and Bi-based oxides broadened the oxidation search space. By the most recent iteration, shown in Figure~\ref{fig:phase}b, the model had successfully predicted the behavior of five new potential barrier materials: Zr, Hf, Ti, Mo, and Ir, with one unsuccessful case: W. The final model shows strong consistency with experiment and captures smooth, interpretable probability gradients across the descriptor space.

\prsec{\label{sec:Discovery of Controlling Process} Controlling process}
The clean separation of data in Figure~\ref{fig:phase} by a single line suggests the existence of a single underlying controlling parameter that determines whether niobium oxide forms. Since our logistic regression model operates in a two-dimensional space of thermodynamic descriptors, the decision boundary it learns corresponds to a linear combination of the interstitial and vacancy formation energies. The line corresponding to a 50\% probability of oxide formation provides the relation
\begin{equation}
    \label{eqn:lr}
    0.56 E(\mathrm{O}_I(\mathrm{M})) - 0.44 E(V_\mathrm{O}(\mathrm{M}_n \mathrm{O}_m)) = -1.36~\text{eV}.
\end{equation}
Physically, this combination can be interpreted as a process that introduces roughly 0.56 of an oxygen interstitial into the metal and fills in 0.44 of a vacancy in the metal oxide. One way to interpret this is as formation of a fraction of a metal oxide unit: oxygen atoms are first inserted into the metal as interstitials, and any remaining vacancies are subsequently eliminated. Quantitatively, this linear combination is reminiscent of an oxide formation process—specifically, the formation of one formula unit of the metal oxide from metal and oxygen atoms normalized by the number of oxygen atoms.

\begin{figure}
    \centering
    \includegraphics[width=\columnwidth]{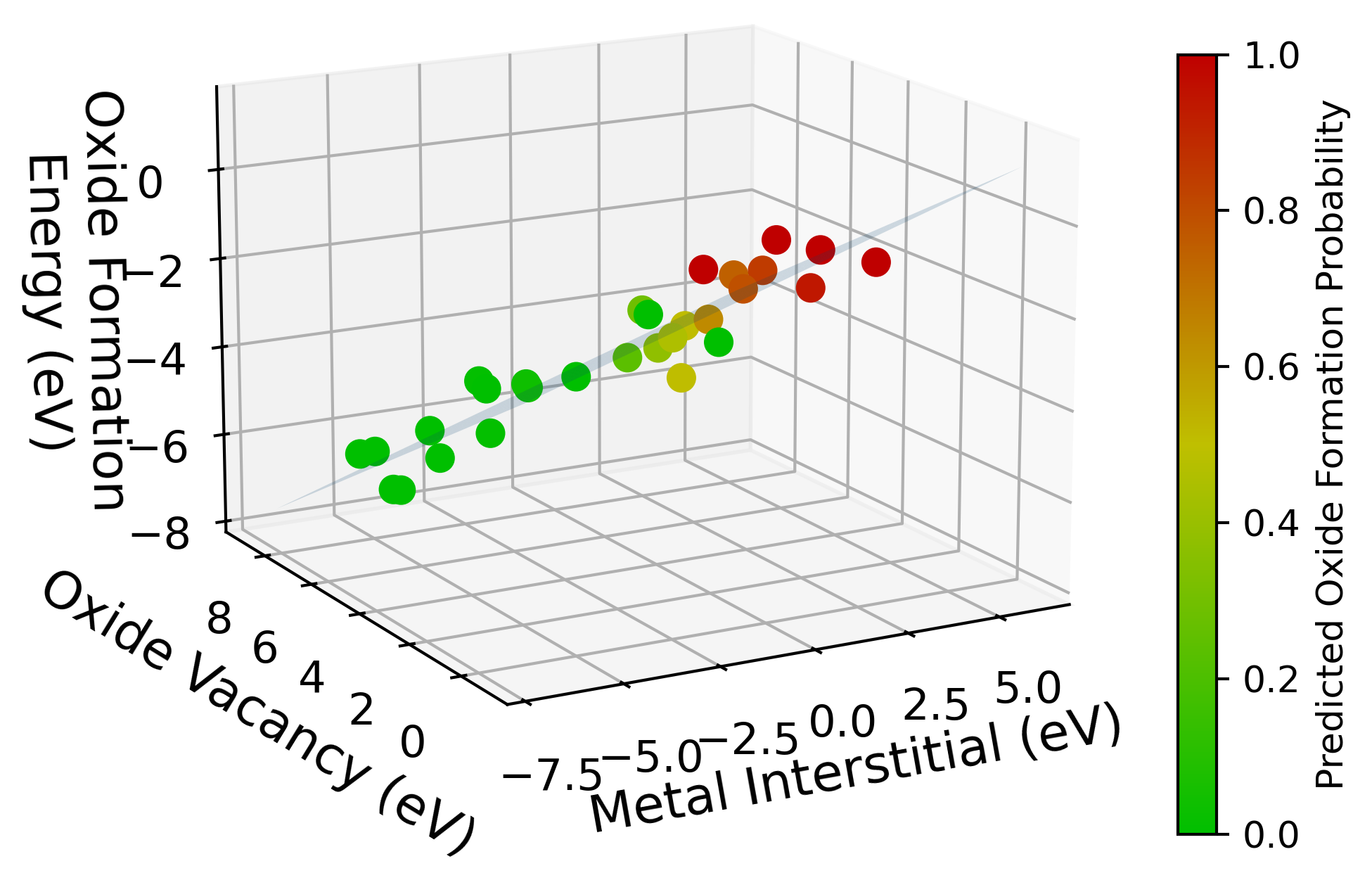}
    \caption{Per-oxygen oxide formation energy plotted against oxygen vacancy and metal interstitial energies for all calculated metals. Point colors indicate predicted oxide formation probabilities from the logistic regression model. A best-fit plane is shown, viewed at an angle that nearly reduces it to a line.}
    \label{fig:fe3d}
\end{figure}

To explore this idea more directly, we examined the formation energy of metal oxides per oxygen atom, $E(\mathrm{M}_n \mathrm{O}_m)/m$ and compared it to our original descriptors. Figure~\ref{fig:fe3d} plots the per-oxygen formation energy against the metal interstitial and oxide vacancy energies for each material that appears in Figure~\ref{fig:phase}b. All materials are found to lie close to a single plane in this three-dimensional space, suggesting a tight linear relationship between these quantities. Accordingly, we fit these data with a least-squares plane, resulting in:
\begin{multline}
    E(\mathrm{M}_n \mathrm{O}_m)/m = (0.19\pm0.05) E(\mathrm{O}_I(\mathrm{M})) \\
    - (0.77\pm0.04) E(V_\mathrm{O}(\mathrm{M}_n \mathrm{O}_m))
\end{multline}
These coefficients correspond to the insertion of approximately 0.19 interstitials and removal of approximately 0.77 vacancies per oxygen atom (for a total insertion of 0.96 atoms), closely aligning with the stoichiometry of a typical metal oxide formula unit normalized to one oxygen atom. This strongly supports the interpretation that our two original descriptors are simply two linear components of the per-oxygen oxide formation energy.

To test whether the per oxygen oxide formation energy is sufficient to explain our logistic regression predictions, we next project a line of constant oxide formation energy (determined by logistical regression holding the slope of the line fixed) back onto the original descriptor space, as seen in Figure~\ref{fig:phase}b. The resulting line separates the data just as well as the decision boundary learned by our original logistic regression model, indicating that per-oxygen oxide formation energy alone is sufficient to reproduce the model’s classification performance.

\begin{figure}
    \centering
    \includegraphics[width=\columnwidth]{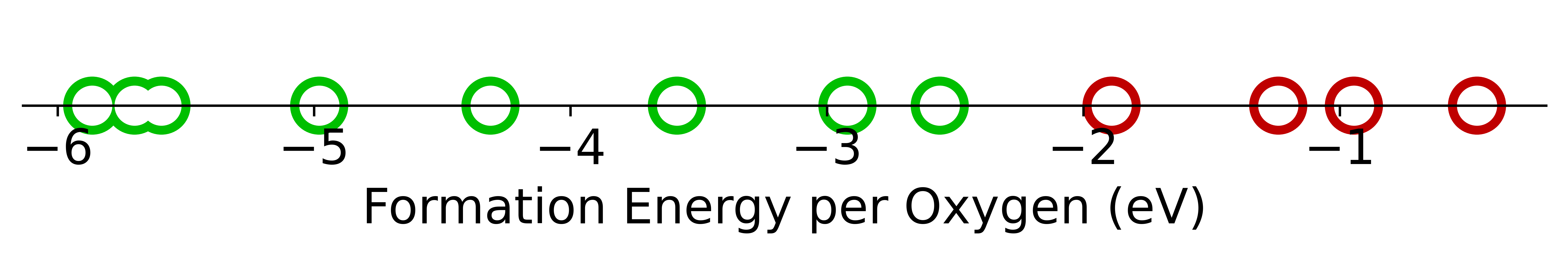}
    \caption{Per-oxygen oxide formation energy of various experimentally tested oxides, colored according to their success (green) or failure (red) to prevent oxidation of the niobium layer.}
    \label{fig:fe}
\end{figure}

Finally, to demonstrate the performance of the per oxygen oxide formation energy as a single descriptor, we plot this quantity for each experimentally tested metal along a single axis (Figure~\ref{fig:fe}), coloring each point according to the observed oxygen barrier performance. A clear separation emerges between effective and ineffective barrier materials. This confirms that oxide formation energy per oxygen atom captures the dominant physics: if forming the oxide is thermodynamically favorable, the metal will tend to prevent oxidation of the buried niobium interface.

\prsec{\label{sec:Optimizing for Coherent Interfaces}Coherent interfaces}
\begin{figure}
    \centering
    \includegraphics[width=\columnwidth]{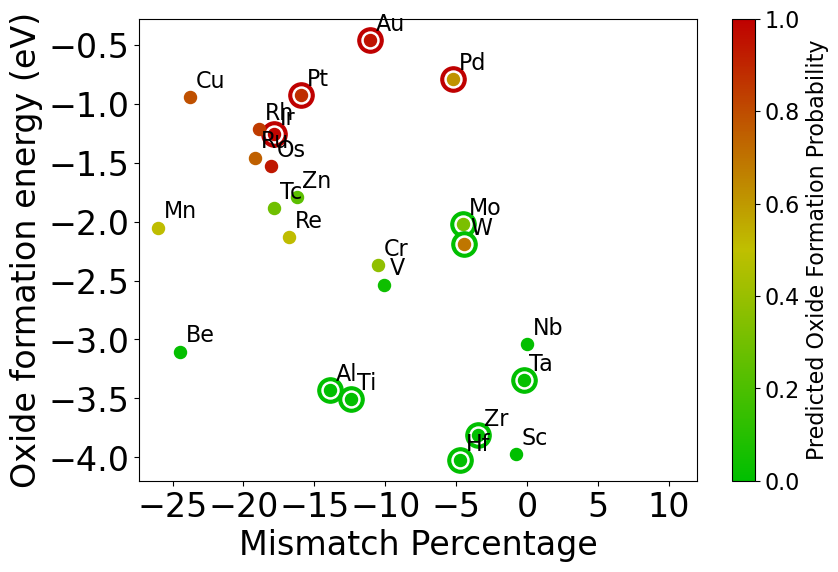}
    \caption{Oxide formation energy versus lattice mismatch percentage against niobium: Metals have been colored according to their logistic predictions, along with outer circles indicating experimental results denoted by their color.}
    \label{fig:lm}
\end{figure}
Having established that oxide formation energy is the primary thermodynamic descriptor governing niobium oxide suppression, we next consider the structural compatibility between candidate barrier metals and niobium. Even materials that thermodynamically inhibit oxide formation may introduce defects if they form incoherent interfaces due to poor lattice matching. Such mismatches can generate dislocations and grain boundaries that act as pathways for oxygen diffusion. To assess this, we computed the lattice mismatch percentage between each candidate metal and BCC niobium, and plotted it against oxide formation energy, as shown in Figure~\ref{fig:lm}.

For face-centered cubic (FCC) metals, the mismatch was calculated using the FCC lattice constant divided by $\sqrt{2}$, corresponding to the in-plane spacing along the interface surface. For hexagonal close-packed (HCP) metals, we used the lattice constant of their lowest-energy FCC polymorph, consistent with the expectation that epitaxial thin films will favor FCC growth over BCC Nb.

We again observe a clear vertical separation between poor and effective oxygen barriers based on oxide formation energy, while the horizontal axis (lattice mismatch) distinguishes good materials from potentially optimal ones. Notably, Hf, Zr, Sc, and Ta exhibit both low oxide formation tendencies and reasonably small mismatch percentages, making them ideal candidates for coherent, oxidation-resistant interfaces. In contrast, metals such as Al and Ti show much higher mismatch values, suggesting they may still be effective as oxygen barriers but could exhibit greater variability or interface instability. From this perspective, we would rank Zr, and potentially Ta, actually as superior to Hf, and Sc (apart from its expense) as a particularly promising candidate that we have not yet explored experimentally. Together, these two material descriptors provide a simple yet powerful framework for identifying barrier materials that balance chemical stability with structural coherence.

\prsec{\label{sec:Conclusion} Conclusion}
This work establishes a closed-loop framework integrating defect-level energetics, logistic-regression modeling, and targeted experiments to guide functional materials design. Applied to superconducting qubit interfaces, it identifies and experimentally validates promising oxygen-barrier materials, including Zr, Ta, and Hf-based capping layers, while highlighting additional candidates such as Sc. Lattice matching further emerges as an important design criterion that refines materials selection, reducing the relative appeal of Hf while elevating Sc despite its cost. More broadly, the approach is applicable wherever physically meaningful, computable descriptors can be defined, enabling accelerated discovery of next-generation materials systems, particularly in experimentally data-sparse regimes limited by the pace of materials synthesis and characterization.

\prsec{\label{sec:Acknowledgments} Acknowledgments}
This work was supported by the Air Force Office of Scientific Research under award number FA9550-23-1-0706. Any opinions, findings, and conclusions or recommendations expressed in this material are those of the author(s) and do not necessarily reflect the views of the United States Air Force. This work made use of the Cornell Center for Materials Research shared instrumentation facility for the XPS analysis system. C.~M\'endez was supported by the US National Science Foundation under award PHY1549132, the Center for Bright Beams.

\prsec{\label{sec:data} Data availability}
The computational data underlying this study — including first-principles inputs and outputs, processed defect energetics, logistic-regression code, and links to the software used to generate the data — are publicly available on GitHub~\cite{Chaudhari2025OxideData}. Experimental XPS measurements used to classify diffusion-barrier performance are available on Zenodo~\cite{zenodo_xps_17665093}.

\bibliography{ref}

\end{document}